\documentclass[prl,aps,twocolumn,showpacs]{revtex4}
\usepackage{graphicx}% Include figure files
\usepackage{textcomp}% for weird symbols
\usepackage[version=3]{mhchem}

\usepackage{graphicx}% Include figure files
\usepackage{dcolumn}% Align table columns on decimal point
\usepackage{bm}% bold math
\usepackage{color}
\begin{document}
\title{Electronic and magnetic properties of zigzag graphene nanoribbons on the (111) surface of Cu, Ag and Au}

\author{Yan Li$^1$}
\author{Wei Zhang$^1$}
\author{Markus Morgenstern$^2$}
\author{Riccardo Mazzarello$^{1}$}\email{mazzarello@physik.rwth-aachen.de}

\affiliation{$^1$ Institute for Theoretical Solid State Physics and JARA, RWTH Aachen University, D-52074 Aachen, Germany\\
$^2$ II. Physikalisches Institut B and JARA-FIT, RWTH Aachen University, D-52074 Aachen, Germany}

\date{\today}

\begin{abstract}
We have carried out an \emph{ab initio} study of the structural, electronic and magnetic properties
of zigzag graphene nanoribbons on Cu$(111)$, Ag$(111)$ and Au$(111)$.
Both, H-free and H-terminated nanoribbons are considered revealing that the nanoribbons invariably possess edge states when deposited on these surfaces.
In spite of this, they do not exhibit a significant magnetization at the edge, with the exception of H-terminated nanoribbons on Au(111), whose
zero-temperature magnetic properties are comparable to those of free-standing nanoribbons.
These results are explained by the different hybridization between the graphene 2p orbitals and those of the substrates
and, for some models, also by the charge transfer between the surface and the nanoribbon.
Interestingly, H-free nanoribbons on Au$(111)$ and Ag$(111)$ exhibit two main peaks in the local density of states around the Fermi energy,
which originate from different states and, thus, do not indicate edge magnetism.

\end{abstract}
\maketitle

Graphene~\cite{Novoselov1} with its remarkable electronic and transport properties~\cite{CastroNeto},
in particular a high room-temperature mobility~\cite{Morozov},
is a promising material for applications in information technology.
While perfect monolayer graphene has a gapless spectrum prohibiting standard transistor applications,
nanostructuring can induce the required band gap.
Recent efforts focused on quasi one-dimensional graphene nanoribbons (GNRs)~\cite{Son,Jiao,Cai,Li}
and zero-dimensional graphene quantum dots (GQDs)~\cite{Ponomarenko,Stampfer,Ritter} and, indeed, revealed a transport gap, e.g.
for GNRs with widths below 10 nm.
Theoretically, unsupported GNRs and GQDs with zigzag edge geometry possess spin polarized edge states
with ferromagnetic order along the edge~\cite{Fujita,Nakada,Wassmann}.
Several experimental studies provide direct~\cite{Ritter,Klusek,Pan,Tao} or indirect~\cite{Chae}
evidence for the presence of the edge states also in supported GNRs, albeit without probing its magnetism.
Such edge states might be exploited for a multitude of spintronics applications~\cite{Son,Kim,Wakabayashi,Wimmer},
but so far it is unclear, if the edge states contribute to the measured transport properties
of nanostructures at all~\cite{Kunstmann}.
Thus, it is important to elucidate the role of the substrate and of edge termination.\\
Recently, several groups including us investigated
in-situ prepared GQDs with exclusive zig-zag-edges,
which are supported by Ir$(111)$~\cite{Pletikosic,Lacovig,Dinesh,Liljeroth,Berndt,Sander,Busse}.
In particular, we revealed the absence of edge states by a combined Density Functional Theory (DFT)
and scanning tunneling microscopy (STM) study~\cite{Yan}.
Our finding was explained by a hybridization of the graphene $\pi$-orbitals with an
Ir 5d$_{z^2}$ surface state at $E_{\rm F}$, which gradually decreases in strength from the edge towards
the center of the GQD and, thus, prohibits a simple shift of the edge state towards the interior of the GQD~\cite{Yan}.
In contrast, STM studies of Tao \emph{et al.}~\cite{Tao} provided convincing evidence for the presence of edge states
in GNRs chemically prepared from carbon nanotubes by so called unzipping~\cite{Dai} and deposited on Au$(111)$.
A peak or a double-peak close to $E_{\rm F}$ was observed in the local density of states (DOS)
close to the edge with a peak distance scaling with the width of the GNR. The double-peak was present for chiral angles
$\theta$ up to 16.1$^\circ$ with respect to the zigzag direction and was ascribed to an antiferromagnetic coupling
between opposite edges~\cite{Tao, Feldner2}, as further evidenced by comparison with results from a Hubbard model Hamiltonian~\cite{Tao}.\\
Here, we present a more realistic description of this system using DFT and including the Au$(111)$ surface.
Large models were used, which enabled us to study GNRs of close to realistic widths and
to correctly describe the lattice mismatch between graphene and Au$(111)$.
For comparison, we also investigated GNRs deposited on the $(111)$ surface of the other two group 1B metals, Cu and Ag.
For all the substrates, we considered both H-free and singly H-terminated GNRs.
%
% chiral angles
%
For H-free GNRs on Au$(111)$, which is most relevant to the experiments by Tao \emph{et al.}~\cite{Tao},
we considered the two chiral angles $\theta = 0^\circ$ and $5^\circ$, while only perfect zigzag GNRs were studied for the other systems.
We show that all the GNRs studied exhibit edge states,
but  that the interaction with the substrate strongly depends on whether the GNR is H-terminated or not and, to a lesser extent, also on the type of substrate.
As a result, it turns out that the edge states are magnetic only when H-terminated and deposited on Au(111). Exactly this system was
prepared experimentally recently~\cite{Dai,Zhang}.
Interestingly, the nonmagnetic H-free GNR on Au$(111)$ also exhibits a double peak around $E_{\rm F}$
(albeit with a much larger splitting as compared to experiments~\cite{Tao}), but the two peaks originate from a hybridization of Au d-orbitals
with the graphene and the unoccupied edge state, respectively, and not from a magnetic splitting of the edge state.\\

%
% Computational methods
%
We considered GNRs with zigzag edges and a width of 8 graphene unit cells.
We employed different supercells in order to account for a) the lattice mismatch between graphene and the $(111)$ surfaces
of Cu, Ag and Au and b) chiral angles.
Here, we describe the models having $\theta=0^\circ$ (the model with finite $\theta$ is discussed in the Supplement~\cite{supplement}).
For Ag and Au a supercell with a size parallel to the GNR six times as large as the unit cell of the $(111)$ surface was used.
For Cu, an extremely large supercell would be required to account for the mismatch, making the calculations unfeasible.
Therefore, a compressed (about 3.8 \%) Cu lattice was instead used to make graphene and Cu(111) commensurate.
A four-layer slab was used for Ag and Au surfaces, whereas the use of a smaller supercell for Cu
(in the direction parallel to the GNR) enabled us to employ thicker slabs containing up to 12 layers.
The latter calculations showed that 4-layer slabs are sufficient to describe
the interaction between the GNR and the Cu surface (see Supplement),
making us confident that the same holds true for Ag and Au substrates.

The structural optimization and the calculation of the electronic and magnetic properties were carried out
using the plane-wave package Quantum-Espresso~\cite{QE}.
We employed gradient-corrected (GGA) exchange correlation functionals~\cite{pbe} and semiempirical
Grimme corrections (DFT-D2) to describe van der Waals interactions~\cite{Grimme}.
Additional computational details are provided in the Supplement~\cite{supplement}.

In the first part of the paper, we present our results about the GNRs on Au$(111)$. Ag and Au substrates are discussed
in the second part and in the Supplement~\cite{supplement}.
Upon DFT relaxation, the H-free GNR with $\theta=0^\circ$ on Au$(111)$
bends considerably: the distance between C atoms and the surface is much shorter
at the GNR edge than in the interior of the ribbon (Fig.~1(a)).
\begin{figure}[t!]
\begin{center}
\includegraphics[width=1.0\columnwidth]{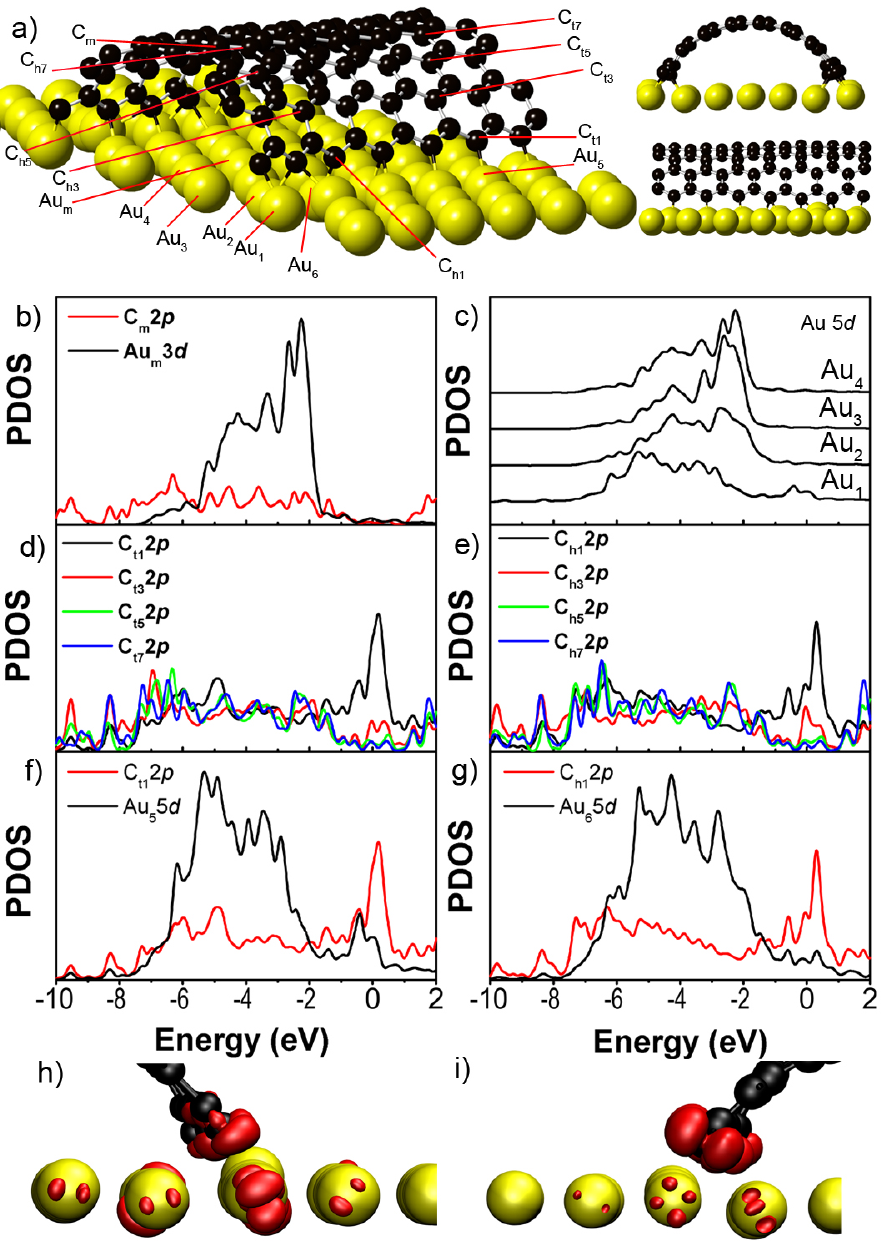}
\end{center}
\caption{
Structural and electronic properties of a H-free GNR on Au$(111)$.
(a) Topography of the relaxed model. For the sake of clarity, only the top Au layer is shown.
    C and Au atoms are labeled by numbers indicating different chemical environments and used in (b)-(g).
    The top and hollow adsorption sites of the edge C atoms are denoted with the subscripts
    ``t'' and ``h'' respectively.
(b) (p$_y$ + p$_z$)-PDOS of a C atom (C$_{\rm m}$) in the middle part of the GNR
    and PDOS of the d states of an Au atom (Au$_{\rm m}$) located beneath.
(c) PDOS of d states of several Au atoms starting from an atom
    below the edge of the GNR (Au$_1$) towards an atom below the centre of the GNR (Au$_{4}$).
(d), (e) (p$_y$ + p$_z$)-PDOS of C atoms at the center of the on-top (d) and hollow (e) region,
    in increasing distance from the edges; row 1 denotes the edge row.
(f), (g) (p$_y$ + p$_z$)-PDOS of an edge C atom at on-top site, C$_{\mathrm{t}1}$ (f),
    and at hollow site, C$_{\mathrm{h}1}$ (g),
    and PDOS of the d states of the nearest neighbor Au atom (Au$_{5}$ and Au$_{6}$ respectively).
(h), (i) Plots of a charge isosurface of a state contributing to the peak
    at -0.2 eV below $E_{\rm F}$ (h) and at 0.3 eV above $E_{\rm F}$ (i).}
\label{PDOS_noH_Au111}
\end{figure}

The maximum distance between C atoms and the Au (as well as Cu and Ag) surface at the center of the GNR and the minimum distance at the edge
are shown in Table I.

\begin{table}
\begin{tabular}{ccc}
\hline \hline                     &           H-free GNR            &        H-terminated GNR        \\
\hline
\;\;\;\;\;\;\;\;\;\;\;\;\;\;\;\;  &   min  \;\;\;\;\;\;\;\;   max   &  min   \;\;\;\;\;\;\;\;   max  \\
\hline
Cu                                &  1.96  \;\;\;\;\;\;\;\;  3.31   &  2.37  \;\;\;\;\;\;\;\;  3.06  \\
Ag                                &  2.18  \;\;\;\;\;\;\;\;  5.61   &  2.69  \;\;\;\;\;\;\;\;  3.16  \\
Au                                &  2.12  \;\;\;\;\;\;\;\;  5.86   &  3.31  \;\;\;\;\;\;\;\;  3.64  \\
\hline \hline
\end{tabular}
\caption{Minimum distance between the edge C atoms of the GNR and the atoms of the relevant (111) surface
and maximum distance between the center of the GNR and the surface. Distances are in Angstrom.}
\label{distances}
\end{table}
A small corrugation (about 0.17 {\AA}) along the GNR edges in accordance with the Moir\'{e} periodicity is observed.
In particular, 4 edge C atoms out of 7 are in a quasi on-top configuration,
whereas the remaining edge atoms are in a quasi bridge or hollow configuration.
In the following, the $z$ axis will be taken perpendicular to the surface and the $x$ axis will be taken parallel to the GNR.
Since the GNR is not parallel to the surface at the edge, the orbitals of C forming edge states are linear combinations of
2p$_z$ and 2p$_y$ orbitals. For the same reason, the dangling-bond orbitals of the edge C atoms of the H-free GNR
(which form $\sigma$ bonds with Au atoms upon deposition on the surface, see below) are also combinations of these orbitals.
%The projected DOS (PDOS) onto 2p and 5d orbitals for different C and Au atoms respectively are shown in Figs.~1(b)-(g).
In Fig.~1(b) the (p$_y$ + p$_z$)-PDOS of a C atom
in the middle part of the GNR and the PDOS of the 5d states of an Au atom located beneath are shown:
the contribution of the PDOS of the d states of Au near $E_{\rm F}$ is very small and the interaction between
the two atoms is basically negligible.
On the other hand, a comparison of the PDOS of the d states of several Au atoms starting from an Au atom
below the edge of the GNR towards an Au atom below the centre (Fig.~1(c)) shows that,
at the edge, the PDOS displays some peaks near $E_{\rm F}$.
Also, the (p$_y$ + p$_z$)-PDOS of the edge C atoms (Figs.~1(d)-(e)) 
exhibit two main peaks near $E_{\rm F}$.
To shed light on the interaction between edge C atoms and Au atoms,
it is useful to compare the respective PDOS at on top sites (Fig.~1(f))
and hollow sites (Fig.~1(g)).
The (p$_y$ + p$_z$)-PDOS of a C atom at an on-top site displays two peaks at -0.2 eV below and 0.3 eV above $E_{\rm F}$ respectively.
Inspection of the PDOS of the nearest neighbor Au 5d orbitals
indicates that a strong hybridization between these states and C states occurs.
 
In particular, a relatively high peak below $E_{\rm F}$ is present, in correspondence with the small
peak of the C p states, whereas a less pronounced peak is found right above $E_{\rm F}$, corresponding to the second,
large peak of the C p PDOS.
In the hollow case, two peaks are present in the PDOS of C p states and Au d states near $E_{\rm F}$ as well. 
The plots of charge isosurfaces of two states contributing to said two peaks in the PDOS of C p orbitals (Figs.~1(h)-(i)) show
that they have a different origin, in that the one above $E_{\rm F}$ is due to the $\pi$ edge states of the GNR,
whereas the one below $E_{\rm F}$ is due to states originating from $\sigma$ bonding between C and Au atoms.
More precisely, the latter peak corresponds to antibonding p-d states, whereas the corresponding bonding states
have much lower energies (about -5 eV below $E_{\rm F}$, see Supplement).
In the on-top case, the 5d$_{z^2}$ and 5d$_{yz}$ orbitals of the Au mostly contribute to the bonding with C p states;
in the hollow case, 5d$_{zx}$ orbitals contribute as well.
Au 6s and 6p orbitals do not play an important role in this bonding.
Since the antibonding $\sigma$ states have lower energies than the edge states,
the latter states are mostly unoccupied and no significant edge magnetism occurs
(less than $3 \cdot 10^{-3} \mu_B$ per edge C atom).

The model of the GNR with $\theta=5^\circ$
also bends considerably and
exhibits essentially non-magnetic edge states for similar reasons (see Supplement).

Next, we consider H-terminated GNRs on Au$(111)$. In this case, the interaction between the GNR and
the surface is weak. There exist two adsorption configurations: in the lower-energy one, the adsorption site of
the edge C atoms changes gradually from on-top to bridge along the edge, and the minimum distance
between the edge C atoms and the surface is 3.31 \AA\hspace{1mm}
(see Table I). The GNR is slightly bent, as evident from Figs.~2(a)-(b).
In the second configuration, the adsorption site of the edge C atoms changes from hollow to (quasi) bridge:
this structure is also magnetic and is discussed in the Supplement.
\begin{figure}[t!]
\begin{center}
\includegraphics[width=1.0\columnwidth]{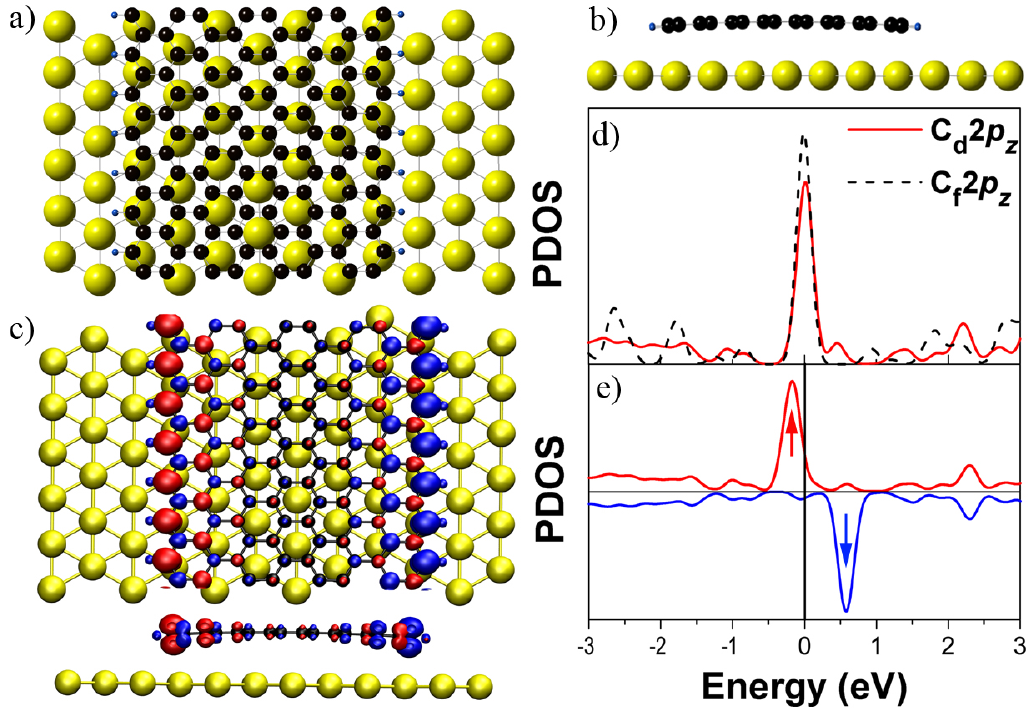}
\end{center}
\caption{
Structural and electronic properties of a H-terminated GNR on Au$(111)$.
(a)-(b) Top and side view of the relaxed model. Only the top Au layer is shown.
(c) Top and side view of an isovalue surface of the edge state spin density of the deposited GNR.
The system has antiferromagnetic order across the GNR.
The red (blue) surface indicates spin up (down) density.
(d) Non-spin-polarized PDOS of the 2p$_z$ orbitals of an edge C atom of the deposited GNR (C$_{\mathrm{d}}$).
The corresponding PDOS of the edge C atom of the free-standing GNR (C$_{\mathrm{f}}$) is also shown for comparison.
(e) Spin-polarized PDOS of the 2p$_z$ orbitals of a C atom at the left edge of the magnetized GNR
shown in Fig.~(c).}
\label{PDOS_H_Au111}
\end{figure}
The electronic properties of the GNR are weakly affected by the presence of the substrate:
the GNR displays a magnetic edge state with antiferromagnetic coupling between the edges
and the magnetization per edge C atom is about 0.22 $\mu _B$
(comparable to the value obtained for free-standing GNRs using equivalent k-point meshes, see Supplement).
An isovalue surface of the edge state spin density is shown in Fig.~2(c), whereas the spin-unpolarized
and spin-polarized PDOS of the 2$p_z$ orbitals of edge C atoms are shown in Figs.~2(d) and (e) respectively.
The energy splitting between spin majority and minority 2p$_z$ peaks on an edge is about 0.7 eV (Fig.~2(e)).

Since the edges of the GNRs investigated experimentally in Ref.~\onlinecite{Tao} exhibit a pronounced
edge curvature, they resemble the H-free case of our calculation.
A comparison of the experimentally observed peak splitting with the calculated one is tempting,
although the experimental GNR width of 8 nm to 20 nm is much larger than in our models (1.6 nm).
 
The experimentally observed energy difference ranges from 50 meV to 20 meV~\cite{Tao} and is an order of magnitude lower than
the one from the DFT calculation.
 
Thus, a direct assignment of the calculated splitting to the experimental results is not possible, maybe due to a different
edge chemistry, but our findings suggest that the interpretation of a double peak around $E_{\rm F}$ alone
as a sign for edge magnetism might be misleading.\\
 
The electronic structure of H-free GNRs on Ag$(111)$ is qualitatively similar to that of the nanoribbons
deposited on Au, as discussed in the Supplement.
On the other hand, H-terminated GNRs on Ag$(111)$ exhibit remarkable differences with respect to Au$(111)$,
which originate from the relatively stronger interaction between the edges of the GNRs and the Ag substrate.
This fact is probably due to the less diffuse character of the 4d orbitals of Ag as compared to the 5d orbitals of Au.
The minimum distance between the C atoms
and Ag$(111)$ is 2.69 \AA\hspace{1mm} at the edge (Table I)
and a more significant bending of the GNR occurs (Figs.~3(a)-(b)).
\begin{figure}[t!]
\begin{center}
\includegraphics[width=1.0\columnwidth]{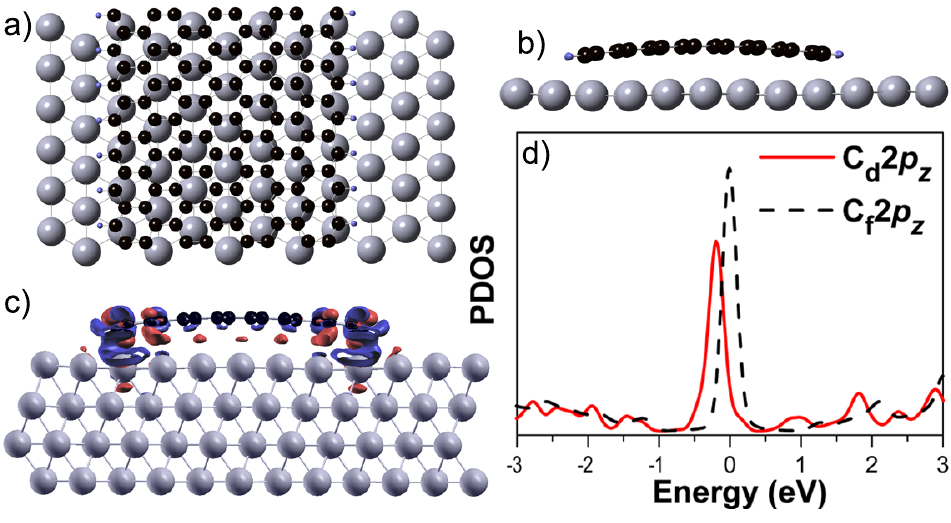}
\end{center}
\caption{
Structural and electronic properties of a H-terminated GNR on Ag$(111)$.
(a)-(b) Top and side view of the relaxed model. Only the top Ag layer is shown.
(c) Isovalue surfaces of the difference between the total charge of the GNR plus substrate system and the charge
of the isolated (bent) GNR and Ag$(111)$.
The red (blue) color indicates accumulation (depletion) of charge.
(d) PDOS of the 2p$_z$ orbitals of an edge C atom of the deposited GNR (C$_{\mathrm{d}}$). The
non-spin-polarized 2p$_z$ PDOS of the edge C atom of the free-standing GNR (C$_{\mathrm{f}}$) is also shown.}
\label{PDOS_H_Ag111}
\end{figure}
 
As a result of the chemical interaction between Ag$(111)$ and the GNR, a significant charge rearrangement
also occurs at the interface, as evident from Fig.~3(c). In total, the GNR has acquired a charge
of $1.6 \cdot 10^{-2}$ electrons per C atom (corresponding to $0.13$ electrons per edge C atom).
This behavior is in contrast to the case of Au$(111)$, where doping is essentially negligible.
The doping of  perfect monolayer graphene has also been shown to depend sensitively
on the type of metal substrate, even in the weak bonding case~\cite{Giovannetti,Vanin,Slawinska,Gebhardt}
(see also the discussion in the Supplement).
 
The PDOS of the C 2p$_z$ orbitals exhibits a large peak (Fig.~3(d)), which corresponds to the edge states.
However, due to the n doping of the GNR, the peak is not exactly at $E_{\rm F}$ but is shifted slightly downwards in energy.
Furthermore, the height of the PDOS peak is reduced with respect to the
free-standing case, due to the interaction with the Ag atoms (Fig.~3(d)).
As a result, no Stoner instability occurs: hence, this system is non magnetic.
In fact, the calculated magnetization is less than $10^{-3} \mu_B$ per edge C atom.
Notice that, in the case of free-standing GNRs, a larger doping density is needed
to fully destroy edge magnetism~\cite{Wimmer,Kunstmann,Jung}.
Therefore, the absence of magnetism in our model is due to a subtle interplay
between charge transfer and the chemical interaction with the substrate.

GNRs on Cu$(111)$ display significant structural differences with respect to the Au and Ag substrate (see Supplement).
In the H-free case, the edge state interacts more strongly with the Cu substrate than with Au$(111)$ and Ag$(111)$:
nevertheless, the PDOS bears a qualitative resemblance to that of the latter models.
Furthermore, the interaction between the GNR and the Cu substrate is not negligible
even in the center of the GNR and the maximum distance between the latter and the surface is only
3.31 \AA\hspace{1mm} (see Table I and Fig.~S2a): the latter property, however, might be due to the use of a compressed surface
or to the van der Waals coefficients employed for Cu (see Supplement) and requires further investigation.
In the H-terminated case, the relatively strong interaction between C and Cu atoms also leads to shorter equilibrium distances
between the GNR and the substrate (Table I and Figs.~S5(a)-(b)).
N-doping leads to the filling of the edge state in this model too, which is also non-magnetic (Figs.~S5(c)-(d)).

In conclusion, our simulations based on DFT indicate that zigzag GNRs deposited on
Cu$(111)$, Ag$(111)$ and Au$(111)$ all possess
edge states but do not exhibit significant edge magnetism, with the exception of H-terminated GNRs on Au$(111)$, whose
zero-temperature magnetization is comparable to that of free-standing GNRs.
These results are explained by the different interaction and charge transfer between the GNRs and the substrates
and show that edge magnetism in zigzag GNRs
%is not very robust and
can be destroyed even upon deposition on a substrate
which interacts weakly with graphene.
Only in the case of H-terminated GNRs on Au$(111)$ is the interaction at the edge sufficiently weak so as
not to affect the electronic and magnetic properties of the edge states significantly.
Hence, our simulations strongly suggest that experimental investigations on edge magnetism in GNRs deposited on metallic
substrates should focus on the latter system.

We acknowledge discussions with  G. Bihlmayer, C. Honerkamp, M. Schmidt and S. Wessel,
the computational resources by the RWTH Rechenzentrum
as well as financial support by DAAD and DFG via Li 1050/2-1 and Mo 858/8-2.

\end{document}

\title{Electronic and magnetic properties of zigzag graphene nanoribbons on the (111) surface of Cu, Ag and Au}%1B metals}

\author{Y. Li$^1$}
\author{Wei Zhang $^1$}
\author{M. Morgenstern$^2$}
\author{R. Mazzarello$^1$}

\email{mazzarello@physik.rwth-aachen.de}

\affiliation{$^1$Institute for Theoretical Solid State Physics and JARA, RWTH Aachen University, D-52074 Aachen, Germany\\
$^2$II. Physikalisches Institut B and JARA, RWTH Aachen University, D-52074 Aachen, Germany}

\date{\today}

\maketitle

In the following supplementary sections, we discuss some computational details (A) and present our
Density Functional Theory (DFT) simulations of
(B) a H-free graphene nanoribbon (GNR) with a finite chiral angle, $\theta = 5^\circ$, deposited on the
Au$(111)$ substrate (consisting of 3 Au layers),
(C) the second adsorption configuration for a H-terminated GNR with $\theta = 0^\circ$ deposited on Au$(111)$,
(D-E) H-free GNRs with $\theta = 0^\circ$ deposited on the Ag$(111)$ (D) and Cu$(111)$ (E) surface (modelled using 4 layer slabs),
(F) H-free GNRs with $\theta = 0^\circ$ deposited on a thick Cu slab containing 12 layers and
(G) H-terminated GNRs with $\theta = 0^\circ$ deposited on Cu$(111)$.
We also discuss doping effects on graphene and GNRs due to metal substrates (H).
%and present results about small models of GNRs on a slightly compressed Au$(111)$ surface (H).
Finally, we also show a charge isosurface of a bonding $\sigma$ state between C atoms at the edge and Au atoms for the case of the
H-free GNR on Au$(111)$ (I).

\section{A. Computational details}

The structural optimization and the calculation of the electronic and magnetic properties were carried out
using the plane-wave package Quantum-Espresso~\cite{QE}.
We employed generalized-gradient (GGA) exchange correlation functionals~\cite{pbe} for all the simulations
(except for some test calculations discussed in section C and H, which were performed within
the local density approximation (LDA)~\cite{pz}).
We used scalar-relativistic ultrasoft pseudopotentials~\cite{Vanderbilt}.
Wave functions were expanded in plane waves with
a kinetic energy cutoff of 30 Ry and a charge-density cutoff of 300 Ry.
The effect of van der Waals (vdW) interactions was described using the semiempirical
correction scheme of Grimme, DFT-D2~\cite{Grimme}.
If one uses the dispersion coefficients C$_6$ and vdW radii R$_0$ provided by Grimme~\cite{Grimme2}
for Cu, Ag and Au, perfect graphene turns out to be n-doped when deposited on the $(111)$ surfaces of these metals.
Experimentally, it has been shown that graphene on Cu$(111)$ and Ag$(111)$ is indeed n-doped~\cite{Shikin},
whereas graphene on Au$(111)$ is p-doped~\cite{Wofford} (see also section H).
Due to this discrepancy, we have computed new vdW parameters for Au by comparing
DFT-D2 and more sophisticated, non-local van der Waals - density functional (vdW-DF)
calculations~\cite{Dion} for graphene on Au$(111)$.
The vdW-DF calculations correctly yield p-doped graphene on Au$(111)$~\cite{Vanin}.
A similar approach had previoulsy been used to determine vdW parameters for Ir~\cite{Busse}.
Although the old and new Au vdW parameters provide different geometries for H-free and H-terminated GNRs on Au$(111)$,
the magnetic properties of both systems do not depend on the set of parameters employed.

It is generally difficult to study d bands using approximate GGA functionals, even when these bands are completely filled
(as occurs for Cu, Ag and Au), in that self-interaction effects can lead to a spurious shift of the bands closer to the Fermi energy.
In spite of this deficiency, we believe that these functionals are able to describe correctly the relatively
strong chemical interaction between the edges of the GNRs and the Cu$(111)$, Ag$(111)$ and Au$(111)$ surfaces.

In the following, the $z$ axis will be taken perpendicular to the metal surface and
the $x$ axis will be taken parallel to the GNR (in the case of chiral angle $\theta=0$),
and the $y$ axis in the plane and normal to the GNR.
To provide a realistic description of the lattice mismatch responsible for the Moir\'{e} pattern,
large supercells must be employed:
since a $(6 \times 6)$ supercell of the primitive cell of Ag$(111)$ and Au$(111)$
corresponds to a good approximation to a $(7 \times 7)$ graphene cell
(resulting in a ratio of the surface unit cell lengths of 1.167,
to be compared with the experimental values of 1.174 and 1.172 respectively),
a $(6 \times 6\sqrt{3})$ supercell of Ag$(111)$ and Au$(111)$ was used for the deposited GNR with $\theta = 0^\circ$.
The supercell parameter of $6\sqrt{3}$ along the $y$ direction, corresponding to a minimum distance of 14.8 {\AA}
between nearest-neighbor periodic images of the (H-terminated) GNR along $y$,
was chosen so as to make the spurious interaction between these images negligible.
Slabs containing 4 Ag (Au) layers were considered and the two topmost Ag (Au) layers
were allowed to relax during structural optimization.
A $2 \times 1 \times 1$ Monkhorst-Pack mesh~\cite{MP}
was employed to perform the integration over the Brillouin zone, which is equivalent
to a $14 \times 1 \times 1$ mesh for the primitive unit cell of the GNR.
Notice that, if one performs a ground-state calculation of a free-standing GNR using
this mesh, one gets a magnetization per edge C atom of 0.22 $\mu _B$,
which is slightly smaller than the fully converged value of 0.27 $\mu _B$ one
obtains using a very dense mesh.
Due to the sheer size of our models, it was not possible to use denser meshes in our simulations.

%
% small models of GNRs
%
%We also performed some simulations of small supercells of H-free and H-terminated GNRs on slightly compressed
%(about 1 \%) Au$(111)$ surfaces, as discussed in section F.

In the case of the H-free GNR on Au$(111)$ with $\theta=5^\circ$,
the employed supercell was twice as large (along the direction of the GNR) as in the $\theta=0^\circ$ case.
For computational convenience, a thinner slab containing three-layers was used in this case.
A $(12\times 6\sqrt{3})$ supercell of Au$(111)$ was employed and
the ground state properties were calculated within the $\Gamma$ point approximation.

For GNRs on Cu$(111)$, a $(1 \times 5\sqrt{3})$ supercell of Cu$(111)$  and a
$14 \times 1 \times 1$ Monkhorst-Pack mesh were employed.
Several slabs with thickness varying from 4 layers to 12 layers were used. 4-layer slabs were found to be sufficient
to correctly model the interaction of the GNR with Cu$(111)$ (at a semi-quantitative level), as discussed in section E.\\
For Au$(111)$ we also performed test calculations of a full graphene layer on a 4-layer Au slab and compared them
with previous results about full graphene layers on 6-layer slabs~\cite{Giovannetti}. We didn't see any significant difference
in the structural and doping properties of graphene (see section H for more details).

For all the models considered, a vacuum layer with thickness in excess of 9 {\AA}
was used to separate the periodic images of the slabs along the $z$ direction.
%For the Cu$(111)$ surface, a value of the unit cell length of 2.742 {\AA} was used.

\section{B. H-free graphene nanoribbon with chiral angle $\theta = 5^\circ$ on the Au$(111)$ surface}

The relaxed model of the H-free GNR with chiral angle $\theta = 5^\circ$ on Au$(111)$ is shown in Fig.~S1(a).
Upon DFT relaxation, this model bends dramatically as well: the maximum distance between the center of the GNR
and the substrate is 5.79 {\AA}, whereas the minimum distance between the edge C atoms and the nearest neighbor Au
atoms is 2.09 {\AA}. Most edge C atoms are in a quasi on-top configuration.
In Fig.~S1(b)-(g) the projected DOS (PDOS) onto the p orbitals (d orbitals respectively) of several C (Au resp.) atoms
having different chemical environment are shown.
Similarly to the case of the GNR with $\theta = 0^\circ$ described in the paper, antibonding $\sigma$ states and
$\pi$ states contribute to the peaks below and above $E_{\rm F}$ respectively and no significant edge magnetism
occurs (about $5 \cdot 10^{-3} \mu_B$ per edge C atom).
%Although the interaction between the GNR and the substrate is more intricate in this case, due to the presence of different
%In Fig.~S1(h)-(i) the states contributing to the peaks in the PDOS of the edge C atoms
%near the Fermi energy are shown.
%
\begin{figure*}[ht!]
\renewcommand\figurename{Fig.~S$\!\!$}
\begin{center}
\includegraphics[width=1.5\columnwidth]{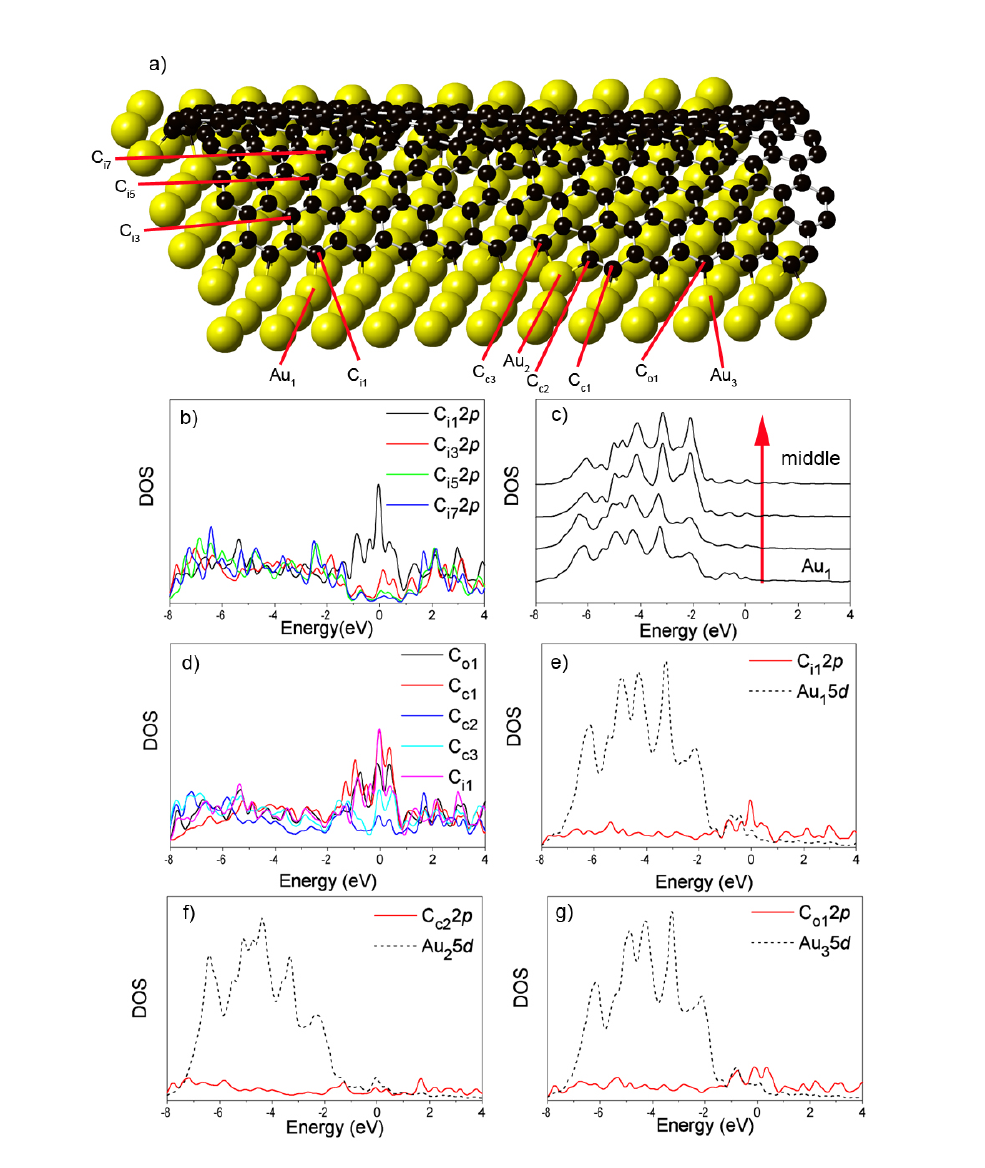}
\end{center}
\caption{Structural and electronic properties of a H-free GNR with chiral angle $\theta = 5^\circ$ deposited on Au$(111)$,
as calculated by DFT.
(a) Topography of the GNR on the Au$(111)$ substrate. For the sake of clarity, only the top Au layer is shown.
    C and Au atoms are labeled by numbers and letters indicating different chemical environments and used in (b)-(g).
(b) PDOS of the d states of several Au atoms starting from an Au atom
    below the edge of the GNR (Au$_1$) towards an Au atom below the centre of the GNR.
(c), (d) Sum of the PDOS of the p$_y$ and p$_z$ orbitals of C atoms in increasing distance from the edges, starting from atom
    C$_{\mathrm{i}1}$ (c) and C$_{\mathrm{o}1}$ (d) respectively. Row 1 denotes the edge row.
(e)-(g) Sum of the PDOS of the p$_y$ and p$_z$ orbitals of 3 C atoms at the edge of the GNR and PDOS of the d states of the
        corresponding nearest-neighbor Au atoms.}
\label{figS1}
\end{figure*}

\section{C. H-terminated graphene nanoribbon on the Au$(111)$ surface: second adsorption configuration and LDA results}

As mentioned in the paper, there are two adsorption configurations for H-terminated GNRs on Au$(111)$ (as well as
Ag$(111)$ and Cu$(111)$).
In the lower-energy one, the adsorption sites of
the edge C atoms change gradually from on-top to bridge along the edge (top-bridge model), whereas
in the second configuration, the adsorption sites change
from hollow to (quasi) bridge (hollow-bridge model), see Figure~S2.
\begin{figure}[ht!]
\renewcommand\figurename{Fig.~S$\!\!$}
\begin{center}
\includegraphics[width=1.0\columnwidth]{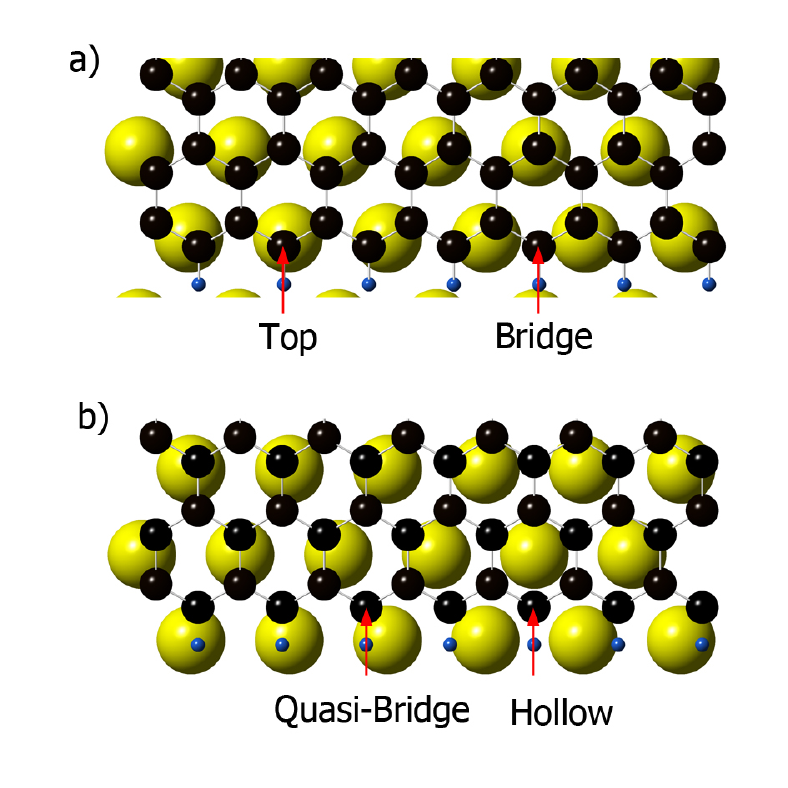}
\end{center}
\caption{Edge structure of the two models of H-terminated GNRs on Au$(111)$.
The C atoms of the two models have different adsorption sites.
(a) Top-bridge configuration (discussed in the paper).
(b) Hollow-bridge configuration, being 6 meV per C atom at the edge higher in energy.}
\label{figS2}
\end{figure}
The energy difference between the two models is about 6 meV per edge C atom.
The magnetization per edge C atom is about 0.22 $\mu _B$ in both models.
In the hollow-bridge model, the minimum distance between edge C atoms and Au atoms is 3.43 {\AA},
whereas the maximum distance between the center of the GNR and the surface is 3.60 {\AA}.\\

We have also investigated these two models using LDA functionals~\cite{pz}. It turns out that,
within this approximation, a) the distance between the GNRs and the substrate is smaller in both configurations
and b) the hollow-bridge model is magnetic, whereas the top-bridge model is non-magnetic.
We believe that the latter result is a spurious result due to the tendency of LDA to overbind.

A second adsorption bridge-hollow configuration exists for H-terminated GNRs on Ag$(111)$ as well.
This configuration is also non-magnetic.

\section{D. H-free graphene nanoribbon on the Ag$(111)$ surface}

The relaxed model of the H-free GNR on Ag$(111)$ is shown in Fig.~S3(a).
As discussed in the paper, the GNR bends significantly upon relaxation.
The distance between the GNR and the substrate at the center of the GNR is slightly smaller
than in the case of Au$(111)$ though (see Table 1 in the paper).
This distance, however, depends sensitively on the vdW parameters employed and could change
if new vdW parameters for Ag were determined using the same approach as for Au.
Very small ripples along the GNR edges in accordance with the Moir\'{e} periodicity are observed.
The maximum height difference between edge atoms is about 0.06 {\AA}.
\begin{figure*}[ht!]
\renewcommand\figurename{Fig.~S$\!\!$}
\begin{center}
\includegraphics[width=1.5\columnwidth]{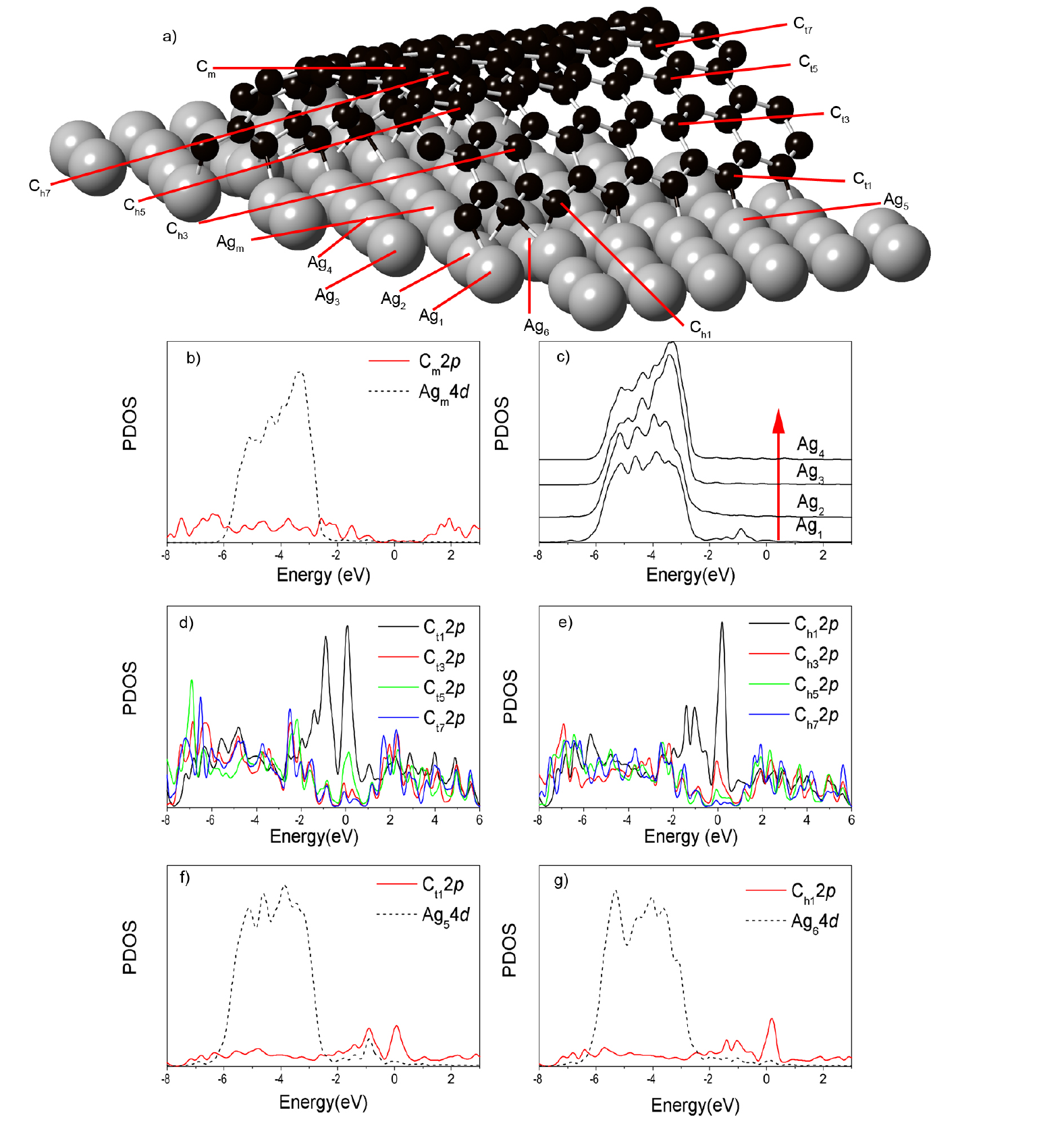}
\end{center}
\caption{Structural and electronic properties of a H-free GNR deposited on Ag$(111)$ as calculated by DFT.
(a) Topography of the GNR on the Ag$(111)$ substrate. For the sake of clarity, only the top Ag layer is shown.
    C and Ag atoms are labeled by numbers and letters indicating different chemical environments and used in (b)-(g).
    The top and hollow adsorption sites of the C atoms at the edge are denoted with the subscripts
    ``t'' and ``h'', respectively.
(b) PDOS onto the p$_y$ and p$_z$ orbitals of a C atom (C$_{\rm m}$) in the middle part of the GNR
    and PDOS of the d states of an Ag atom (Ag$_{\rm m}$) located beneath.
(c) PDOS of the d states of several Ag atoms starting from an Ag atom
    below the edge of the GNR (Ag$_1$) towards an Ag atom below the centre of the GNR (Ag$_{4}$).
(d), (e) Sum of the PDOS of the p$_y$ and p$_z$ orbitals of C atoms at the center of the on-top (d) and hollow (e) region,
    in increasing distance from the edges; row 1 denotes the edge row.
(f) Sum of the PDOS of the p$_y$ and p$_z$ orbitals of a C atom at the edge of the GNR at on-top adsorption site (C$_{\mathrm{t}1}$)
    and PDOS of the d states of the Ag atom beneath (Ag$_{5}$).
(g) Sum of the PDOS of the p$_y$ and p$_z$ orbitals of a C atom at the edge of the GNR at hollow adsorption site (C$_{\mathrm{h}1}$)
    and PDOS of the d states of a nearest neighbor Ag atom (Ag$_{6}$).}
%(h), (i) Plots of a charge isosurface of a state contributing to the peak
%    at ? eV below $E_{\rm F}$ (h) and at ? eV above $E_{\rm F}$ (i).}
\label{figS3}
\end{figure*}
In Figs.~S3(b)$-$(g), the projected DOS (PDOS) onto the 2p$_y$ and 2p$_z$ orbitals for different C atoms
are shown in combination with the PDOS of the 4d orbitals of neighboring Ag atoms.
As occurs in the case of H-free GNRs on Au$(111)$, the edges of the GNR can be roughly divided into two regions,
in which C atoms sit at quasi on-top sites and at bridge or hollow sites respectively.
C atoms at the edge row hybridize strongly with Ag $d$ orbitals and form bonding and antibonding $\sigma$ states,
which are both occupied. The antibonding state around $E=-1.0$ eV can be nicely seen in Figs.~S3(f) and (g).
The interaction of the $\pi$ edge state with the surface is weaker (Figs.~S3(f) and (g)) and the state is mostly localized at the
1st C edge row (Figs.~S3(c) and (d)). Moreover, the PDOS peak corresponding to this state is about 0.15 eV
above $E_{\rm F}$ and the state is thus non-magnetic.

\section{E. H-free graphene nanoribbons on the Cu$(111)$ surface}

The relaxed model of the H-free GNR on Cu$(111)$ is shown in Fig.~S4(a).
Due to the stronger interaction between the GNR and the Cu substrate with respect to the Au and Ag surfaces,
the distance between the GNR and Cu$(111)$ (in particular, at the center of the GNR)
is relatively small (Table~1 in the paper).
Therefore, although the GNR bends significantly upon relaxation, the curvature is not quite as large as for
Au$(111)$ and Ag$(111)$ substrates.
As already mentioned in section D, the distance between the GNR and the substrate at the center of the GNR
depends sensitively on the vdW parameters used. This issue certainly deserves further investigation.
if new vdW parameters for Ag were determined using the same approach as for Au.
Moreover, in this model the two edges of the GNR are not equivalent, in that the C atoms at the two edges have
different positions with respect to the substrate. Nevertheless, the interaction with the substrate is strong
for both edges and the GNR is non-magnetic.
\begin{figure*}[ht!]
\renewcommand\figurename{Fig.~S$\!\!$}
\begin{center}
\includegraphics[width=1.3\columnwidth]{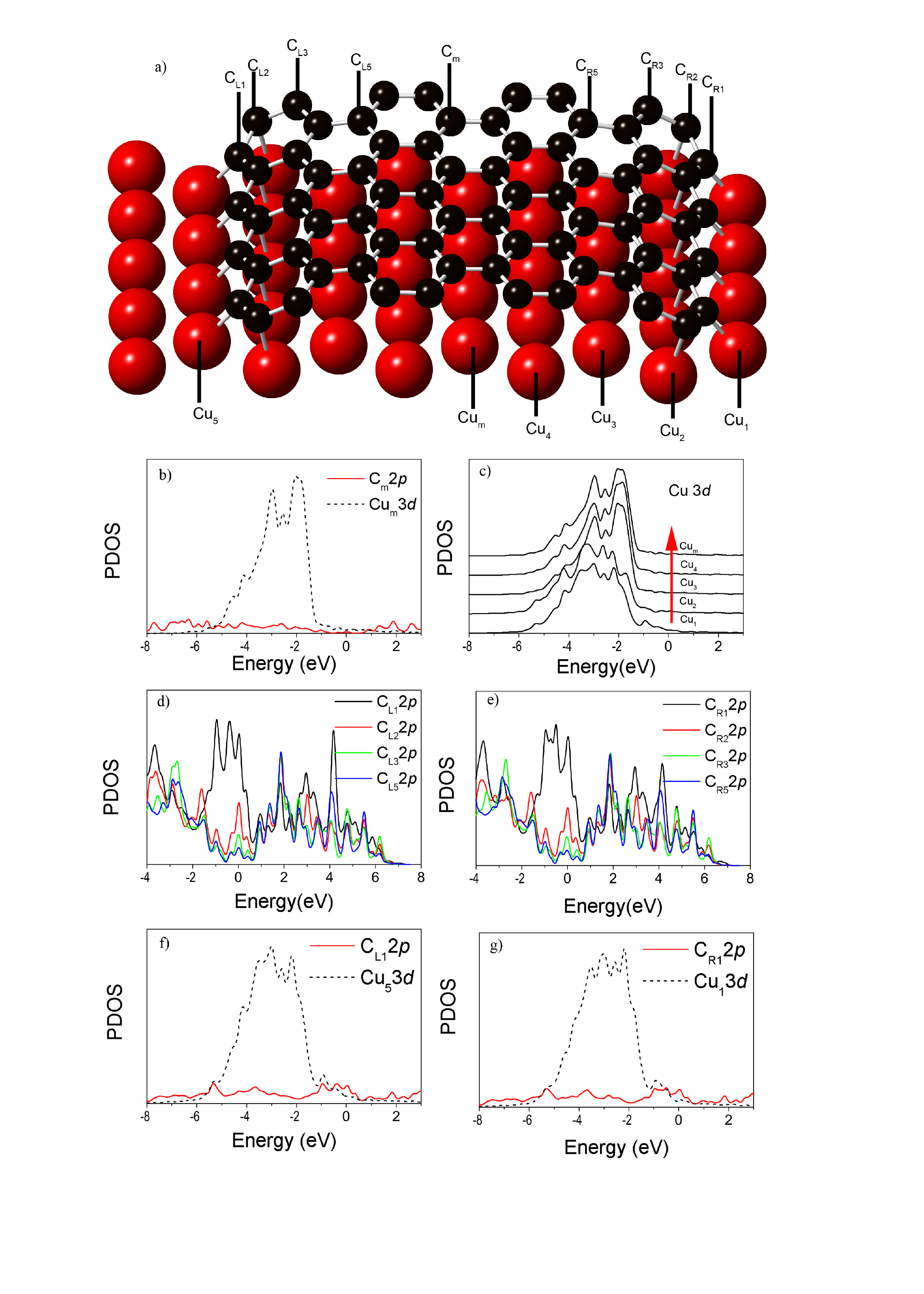}
\end{center}
\caption{Structural and electronic properties of the H-free GNR deposited on Cu$(111)$ as calculated by DFT.
(a) Topography of the GNR on the Cu$(111)$ substrate. For the sake of clarity, only the top Cu layer is shown.
    C and Cu atoms are labeled by numbers indicating different chemical environments and used in (b)-(g).
    The two edges are not equivalent, in that the C atoms at the two edges have different positions with respect
    to the substrate: for this reason, the PDOS of atoms at both
    left (subscript ``L'') and right (subscript ``R'') edge are shown.
(b) PDOS onto the p$_y$ and p$_z$ orbitals of a C atom (C$_{\rm m}$) in the middle part of the GNR
    and PDOS of the d states of a Cu atom (Cu$_{\rm m}$) located beneath.
(c) PDOS of the d states of several Cu atoms starting from a Cu atom
    below the edge of the GNR (Cu$_1$) towards a Cu atom below the centre of the GNR (Cu$_{\rm m}$).
(d)-(e) Sum of the PDOS of the p$_y$ and p$_z$ orbitals of C atoms in increasing distance from the
    left (d) and right (e) edge; in both figures, row 1 denotes the relevant edge row.
(f) Sum of the PDOS of the p$_y$ and p$_z$ orbitals of a C atom at the left edge of the GNR (C$_{\mathrm{L}1}$)
    and PDOS of the d states of the nearest-neighbor Cu atom (Cu$_{5}$).
(g) Sum of the PDOS of the p$_y$ and p$_z$ orbitals of a C atom at the right edge of the GNR (C$_{\mathrm{R}1}$)
    and PDOS of the d states of the nearest-neighbor Cu atom (Cu$_{1}$).}
%(h)-(i) Sum of the PDOS of the p$_y$ and p$_z$ orbitals of a C atom at the left (h) and right (i) row 2
%    and PDOS of the d states of the corresponding nearest-neighbor Cu atom.}
\label{figS4}
\end{figure*}
In Figs.~S4(b)$-$(g), the projected DOS (PDOS) onto the 2p$_y$ and 2p$_z$ orbitals for different C atoms
are shown, as well as the PDOS of the 3d orbitals of several Cu atoms.
Similarly to the case of H-free GNRs on Cu$(111)$ and Ag$(111)$, C atoms at the edge row hybridize with Cu $d$ orbitals
and form bonding and antibonding $\sigma$ states, which are both occupied,
as the corresponding PDOS shows (Fig.~S4(f) and (g)).
The interaction of the $\pi$ edge state with the surface is relatively weak (albeit stronger than in the case of
Au$(111)$ and Ag$(111)$ substrates), and the state is mostly localized at the
1st C edge row. Moreover, the PDOS peak corresponding to this state is about 60 meV
above $E_{\rm F}$ and the state is non-magnetic (Figs.~S4(f) and (g)).

\section{F. H-free graphene nanoribbons on a 12-layer Cu slab}

In this section, we briefly discuss our simulations of a H-free GNR on Cu$(111)$, wherein the surface was modeled
using a thick slab containing 12 Cu layers. We compare the results of these simulations with those obtained by
using a four-layer slab. Although thick slabs are generally needed to describe quantitatively some properties of
surfaces, such as the dispersion of surface states, it turns out that 4-layer slabs are already sufficient to describe
the interaction between the GNR and the Cu surface.
In particular, the PDOS of the C atoms at the edge and the PDOS of the nearest neighbor Cu atoms are very similar in the
two cases, as shown in Figs.~S5(a)-(b). These results make us confident that 4-layer slabs are sufficient to describe
the interaction between GNRs and the Ag$(111)$ and Au$(111)$ surfaces as well.
\begin{figure*}[ht!]
\renewcommand\figurename{Fig.~S$\!\!$}
\begin{center}
\includegraphics[width=2.0\columnwidth]{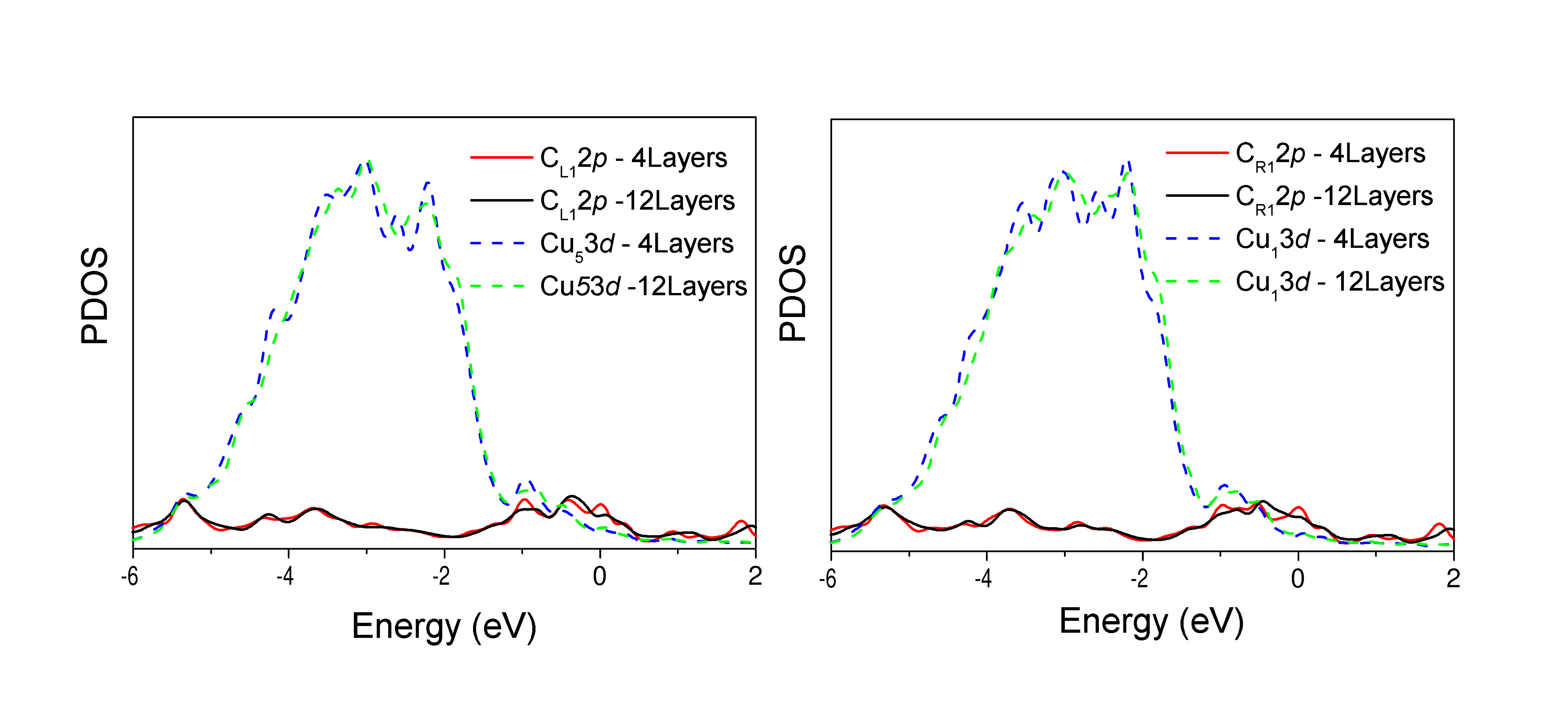}
\end{center}
\caption{
(a) Sum of the PDOS of the p$_y$ and p$_z$ orbitals of a C atom at the left edge of the GNR shown in Figure~S4
    and PDOS of the d states of a nearest-neighbor Cu atom, calculated for the two models having a 4-layer and
    a 12-layer Cu substrate respectively. Notations are the same as in Figure~S4.
(b) Sum of the PDOS of the p$_y$ and p$_z$ orbitals of a C atom at the right edge of the GNR in Figure~S4
    and PDOS of the d states of a nearest-neighbor Cu atom, calculated for the same two models.
    Notations are the same as in Figure~S4.}
\label{figS5}
\end{figure*}

\section{G. H-terminated graphene nanoribbons on the Cu$(111)$ surface}

The relaxed model of the H-terminated GNR on Cu$(111)$ is shown in Figs.~S6(a)-(b).
The strong interaction between C and Cu atoms leads to relatively short equilibrium distances between the
GNR and the substrate in the H-terminated case as well (see Table~1 in the paper).
\begin{figure*}[ht!]
\renewcommand\figurename{Fig.~S$\!\!$}
\begin{center}
\includegraphics[width=2.0\columnwidth]{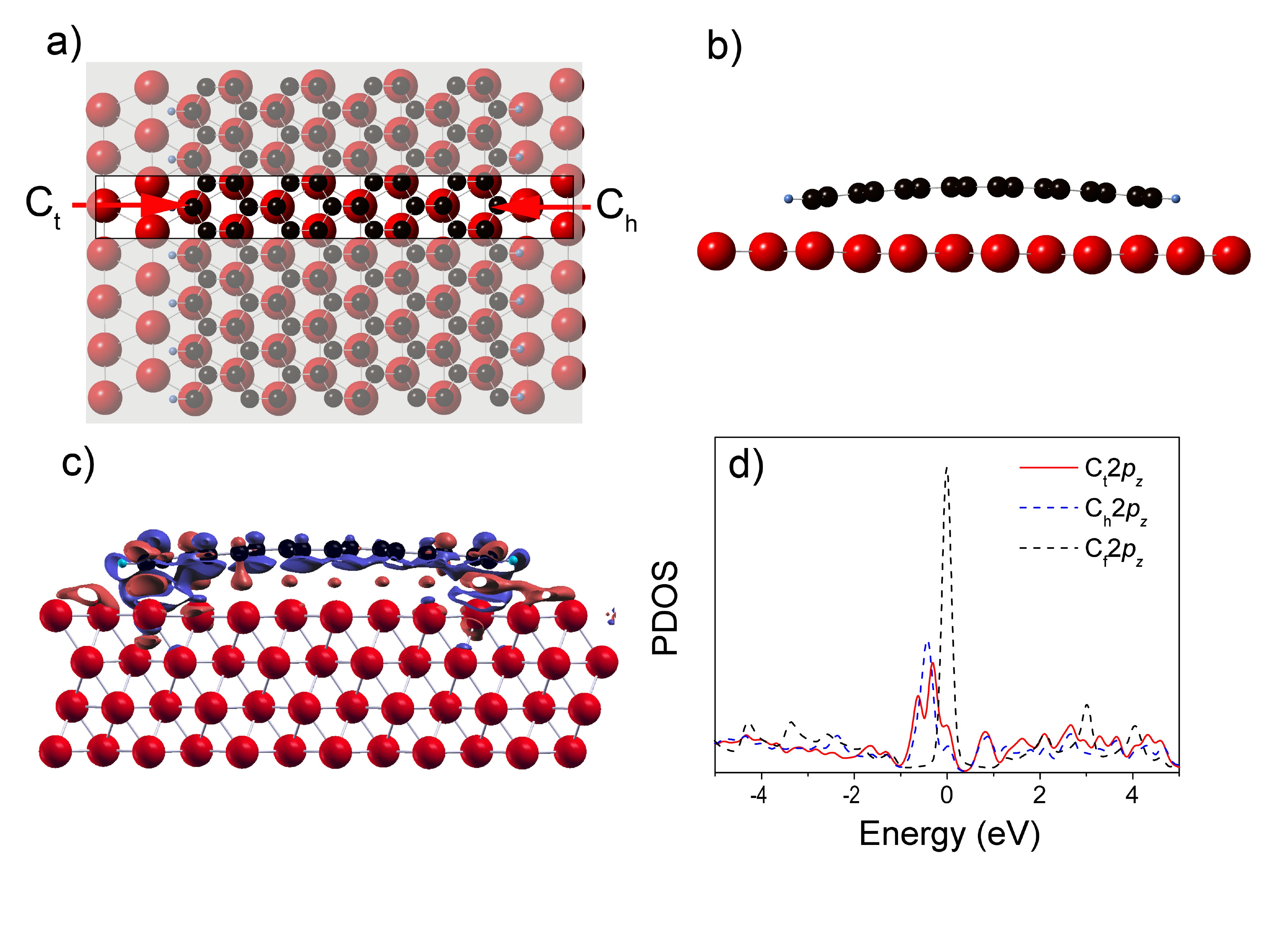}
\end{center}
\caption{
Structural and electronic properties of a H-terminated GNR on Cu$(111)$.
(a)-(b) Top and side view of the relaxed model. Only the top Cu layer is shown.
The C atoms at the two edges have different adsorption sites,
top (C$_{\mathrm{t}}$) and hollow (C$_{\mathrm{h}}$) respectively.
(c) Isovalue surfaces of the difference between the total charge of the GNR plus substrate system and the charge
of the isolated (bent) GNR and Cu$(111)$.
The red (blue) color indicates accumulation (depletion) of charge.
(d) PDOS of the 2p$_z$ orbitals of the edge C atoms of the deposited GNR.
The non-spin-polarized 2p$_z$ PDOS of the edge C atom of the free-standing GNR (C$_{\mathrm{f}}$)
is also shown.}
\label{figS6}
\end{figure*}
The charge transfer from the substrate to the GNR, of the order of $2 \cdot 10^{-2}$ electrons per C atoms,
is slightly more significant than for Ag$(111)$ and shifts the edge state further down in energy (Figs.~S6(c)-(d)).
As for the Ag$(111)$ substrate, the calculated magnetization is less than $10^{-3} \mu_B$ per edge C atom.
It is interesting to note that, in our relaxed model, the C atoms at the two edges have different adsorption sites,
top and hollow respectively (Fig.~S6(a)).
The two adsorption sites correspond to the adsorption configurations (top-bridge and hollow-bridge)
discussed for GNRs on Au$(111)$. In the case of the model of GNR on Cu$(111)$, however, there is only one type of adsorption site
for each configuration, due to the absence of a lattice mismatch.
At the ontop configuration, the edge state hybridizes more strongly
with the d orbitals of the neighboring Cu atoms than in the hollow case: as a result, the corresponding peak in the PDOS
of the C 2p orbitals is broader (Fig.~S6(d)).

\section{H. Doping of graphene nanoribbons due to the metal substrate}

In a recent DFT work~\cite{Giovannetti} it was shown that graphene doping due to a metal substrate
originates not only from electron transfer between the metal and the graphene levels (so as to match the
work functions of the metal and graphene)
but also from the metal-graphene repulsive chemical interaction.
In other words, the doping level was shown to be affected by a potential step at the interface, which
stems from a redistribution of the charge due to Pauli repulsion (so called "pillow effect")~\cite{Gebhardt}.
As a result, at the equilibrium separation between graphene and the substrate,
the transition from n-type to p-type doping of graphene does not occur for
W$_g$ = W$_s$ (where W$_g$ and W$_s$ are the work functions of free graphene and the clean substrate, respectively)
but for W$_s$ $>$ W$_g$.
Using the local-density approximation (LDA), Giovannetti {\it et al.} determined a critical value
of W$_s$ $-$ W$_g$ $\sim$ 0.9 eV~\cite{Giovannetti}.

As already mentioned in section A, graphene is n-doped when it is deposited on Cu$(111)$ and Ag$(111)$~\cite{Shikin},
whereas it is slightly p-doped when deposited on Au$(111)$~\cite{Wofford}.
Our GGA+vdW calculations correctly reproduce this behavior for perfect graphene (in the case of Au, new
vdW parameters had to be computed to obtain the correct doping, see section A).
As regards H-terminated GNRs, the behavior of GNRs on Cu$(111)$ and Ag$(111)$ surfaces is in qualitative agreement
with that of bulk graphene, however we find no significant doping in our simulations
of H-terminated GNRs on Au$(111)$.
The reason for this difference
%between bulk graphene and narrow GNRs
is the fact that, in general, the chemical interaction
at the edges of a deposited zigzag GNR is quite different from that of deposited graphene, due to the presence of the edge states,
and can change the doping of narrow GNRs with respect to graphene.

We would also like to stress that the use of thin slabs containing 4 layers does not affect the doping
character of graphene, nor its distance from the substrate.
We repeated some of the calculations performed in Ref.~\onlinecite{Giovannetti}
(i.e. we investigated the electronic properties of a monolayer of graphene on Au$(111)$
using small supercells and LDA functionals), employing 4-layer slabs instead of
6-layer ones (which they employed), and found doping levels in good agreement (within 3 \%) with theirs.
Also, the difference in the distance between graphene and Au$(111)$ was less than 0.01 {\AA}.
This is yet another test calculation which shows that 4-layer slabs are sufficient to describe accurately
the interaction between graphene and the metal surfaces.

\section{I. Bonding p-d states between edge C atoms and Au atoms}

In Fig.~S7 we show the plot of a charge isosurface of a state of the H-free GNR on Au$(111)$
at about $-5$ eV below the Fermi level. At the edge, this state is made
of C 2p orbitals and 5d orbitals of the nearest neighbor Au atoms and has a bonding character.
Hence, it is the partner of the antibonding state shown in Fig.~1 of the main text.
\begin{figure*}[ht!]
\renewcommand\figurename{Fig.~S$\!\!$}
\begin{center}
\includegraphics[width=1.0\columnwidth]{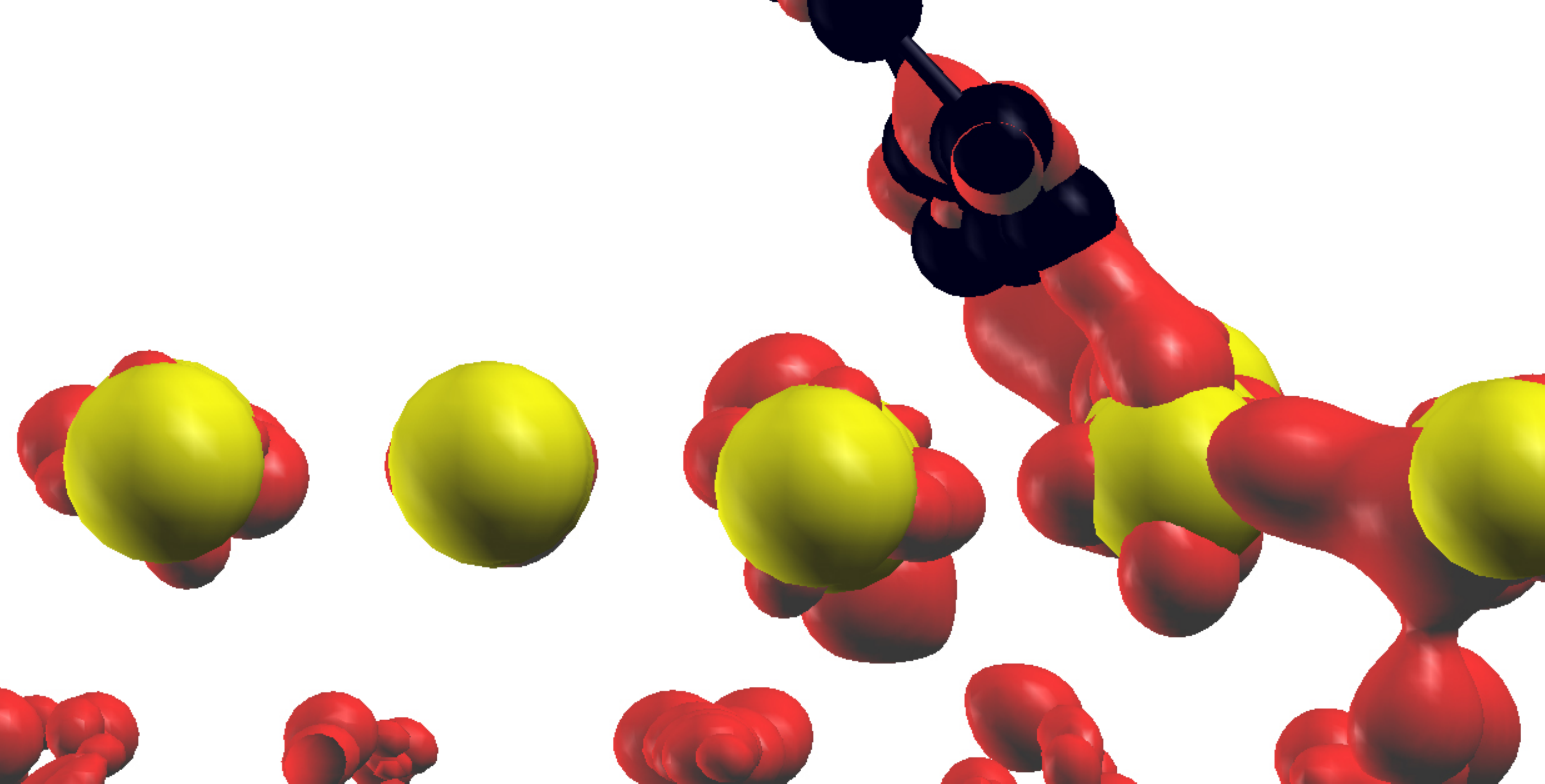}
\end{center}
\caption{Plot of a charge isosurface of a p-d bonding state of the H-free GNR on Au $(111)$
at -5 eV below $E_{\rm F}$. This state is the partner of the antibonding state shown in Fig.~1 of the paper.}
\label{figS7}
\end{figure*}